\documentclass[letterpaper,10 pt,conference]{ieeeconf}
\IEEEoverridecommandlockouts
\usepackage{amsmath}
\usepackage{amssymb}
\usepackage{booktabs}
\usepackage{graphicx}
\usepackage{mathptmx}
\usepackage{subfig}
\usepackage{xfrac}
\usepackage[noend]{algorithmic}
\usepackage[ruled,lined]{algorithm2e}

\title{\LARGE \bf Path Planning and Task Assignment for Data Retrieval from Wireless Sensor Nodes Relying on Game-Theoretic Learning}

\author{
	Sotiris Papatheodorou,~Michalis Smyrnakis,~Tembine Hamidou and Anthony Tzes
	\thanks{The authors are with New York University Abu Dhabi, Electrical and Computer Engineering, Abu Dhabi, P.O.Box 129188, United Arab Emirates}
	\thanks{Corresponding author's e-mail: sotiris.papatheodorou@nyu.edu}
	\thanks{This research work was partially supported by the U.S. Air Force Office of Scientific Research under grant number FA9550-17-1-0259 and by the European Union Horizon 2020 Research and Innovation Programme under the Grant Agreement No. 644128, AEROWORKS.}
}

\begin{document}
\maketitle
\thispagestyle{empty}
\pagestyle{empty}

\begin{abstract}
	The energy-efficient trip allocation of mobile robots employing differential drives for data retrieval from stationary sensor locations is the scope of this article.
	Given a team of robots and a set of targets (wireless sensor nodes), the planner computes all possible tours that each robot can make if it needs to visit a part of or the entire set of targets.
	Each segment of the tour relies on a minimum energy path planning algorithm.
	After the computation of all possible tour-segments, a utility function penalizing the overall energy consumption is formed.
	Rather than relying on the NP-hard Mobile Element Scheduling (MES) MILP problem, an approach using elements from game theory is employed.
	The suggested approach converges fast for most practical reasons thus allowing its utilization in near real time applications.
	Simulations are offered to highlight the efficiency of the developed algorithm.
\end{abstract}


\section{Introduction}

A typical Wireless Sensor Network (WSN) \cite{Potdar_2009WAINA} consists of a
number of low power, static sensor nodes scattered in the environment and a
base station. The nodes need to transfer the data collected by their sensors
back to the base station in an energy efficient manner. Given the low
transmission power of the nodes, a WSN consisting solely of static nodes would
require a large number of densely placed relay nodes in order to transmit the
data to the base station. This approach however significantly increases the
cost of the WSN and makes large scale deployment more difficult.

In order to counteract this problem, the usage of mobile robots as data mules
has been proposed \cite{Bhadauria_2011JFR,Tsilomitrou_2017MED}. The robots
download data from one or more sensor nodes and upload it to the base station.
Since the sensor nodes have a limited amount of on-board memory, the data mule
robots must visit each node in a timely fashion which is called the Mobile
Element Scheduling (MES) problem \cite{Somasundara_2004RTSS}. The problem of
planning a route for each data mule robot is an NP-hard optimization problem
which is formulated as the Traveling Salesman Subset-tour Problem (TSSP) \cite{Applegate_2011book}.

Moreover, the limited battery capacity of the mobile robots has to be taken
into account when planning the routes of the mobile robots. To that extent an
energy model of the mobile robots is required. Work has been done in both the
energy modeling \cite{Wahab_2015SoutheastCon} of various types of ground
robots as well as energy efficient path planning for them
\cite{Mei_2004ICRA,Tokekar_2014AR}.

In this article we assume a team of car-like robots which act as data mules in
a WSN. Our goal is to create energy efficient paths so that the team of mobile
robots collectively visits all sensor nodes. In order to avoid the
computationally expensive Mixed-Integer Linear Programming (MILP) problem
associated with finding the optimal solution to the TSSP, we formulate the
problem as a cooperative game and employ game theoretic learning algorithms.
Although these algorithms converge to suboptimal solutions, they are typically
magnitudes faster compared to MILP. Thus the main contribution of this article
is the utilization of game-theoretic learning techniques in order to reduce the
computational burden of routing the data mules, while also providing energy
efficient trajectories by incorporating path energy costs into the game utility
function.

Section \ref{sec:problem_statement} presents the particular problem under
consideration and outlines the way it is approached. Section \ref{sec:energy}
presents the agent model as well as the derivation of energy efficient paths
and velocity profiles based on that model. Section \ref{background}
provides some theoretical background on the game-theoretic methods and
algorithms used. Section \ref{sec:utility} defines the particular utility
function defined for the problem. Section \ref{sec:simulations} provides
several comparative simulation studies between various game-theoretic learning
algorithms and is followed by concluding remarks.

\section{Problem statement}
\label{sec:problem_statement}
We assume $M$ sensor nodes comprising a WSN and dispersed in a convex region
$\Omega \subset \mathbb{R}^2$ at positions $P_j \in \Omega, \ j \in I_{M}$,
where $I_{M} = \left\{1, \ \dots, \ M \right\}$. We also assume a team of $N$
mobile robots with initial poses
$Q_i^0 = \left[{q_i^0}^T \ \theta_i^0 \right] \in \Omega \times \left[-\pi, \pi\right] $
which act as data mules for the sensor nodes. The goal of the robot team is to
collectively visit all sensor nodes and return to its initial position, while
also reducing the total energy consumption as much as possible.

In order to formulate the aforementioned problem as a game, we need to define
the action space of each robot agent. The action of each agent consists of a
sequence of sensor nodes, which essentially defines a route for the agent.
However, as the size of the WSN increases, the quick growth of the action space
of each agent makes the problem infeasible computationally. In order to combat
this problem, we use two techniques outlined in the sequel.

First, by imposing a maximum number $M^{\max}$ of sensor nodes any single agent
may visit during a single route, we can effectively limit the size of the
action space. This constraint makes sense from a practical point of view as
well, since a single agent may not have a large enough energy capacity in order
to visit a large number of nodes without the need to recharge. Using the upper
limit in the number of nodes to visit, the number of available actions $A_i$
for a single agent is:
\begin{equation*}
	A_i = 1 + \sum_{k=1}^{M^{\max}} \frac{M!}{(M-k)!},
\end{equation*}
where one is added in order to include the case of the agent visiting zero
sensor nodes.

In addition, especially in the case of large WSNs, it does not make sense for
some agent $i$ to cover a large distance in order to visit a sensor node $j$
which is significantly closer to one or more other agents. More importantly,
the robot agent need to communicate in order to choose their routes in a
cooperative manner, but due to their finite communication range, only agents in
relative proximity will be able to do so. To that extent, we split the robot
team into groups consisting of agents whose communication graph is fully
connected. After the robot team has been split into sub-teams, each sub-team
can be assigned a region of responsibility and be responsible for visiting only
the sensor nodes inside this region. One of the possible methods for assigning
regions of responsibility is using the Generalized Voronoi diagram
\cite{Okabe_1994IJGIS}.

\begin{figure}
	\centering
	\includegraphics[width=0.25\textwidth]{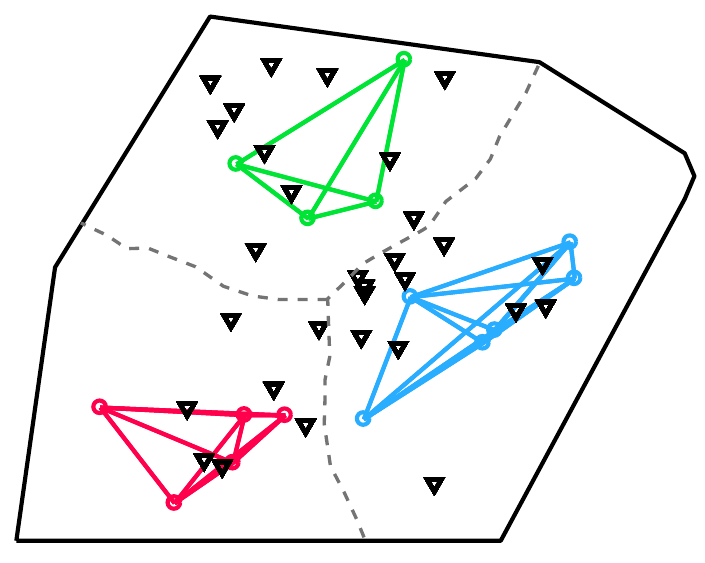}
	\caption{Splitting of the robot team into subgroups and region assignment
	using the Generalized Voronoi diagram.}
	\label{fig:problem_statement}
\end{figure}

The Generalized Voronoi diagram for a set of $n$ planar sites $H_i, \ i \in
I_N$ inside some planar region $\Omega$ is defined as:
\begin{align*}
	V_i = \left\{q \in \Omega \colon d\left(q,H_i\right) \leq
	d\left(q,H_j\right), \ \forall j \neq i
	\right\}, \ i \in I_N
\end{align*}
where $d\left(q,D\right) = \min \left\{\parallel q - p \parallel, \forall p \in
D \right\}$. By computing the convex hull of the robot positions of each group,
we can represent each group by a convex polygon. We can then compute the
Generalized Voronoi diagram using these polygons as seeds and thus assigning a
region of responsibility to each group. This procedure is illustrated in Figure
\ref{fig:problem_statement}, where agent groups and their communication links
are shown in colored lines, the sensor nodes are marked with black triangles
and the boundaries of the Generalized Voronoi diagram are shown in dashed gray.

\section{Energy efficient path computation}
\label{sec:energy}
\subsection{Agent Kinematic Model}
\label{sec:model}
In order to compute energy efficient paths, an appropriate agent model is
required. The agent model used is the same car-like model as the one in
\cite{Tokekar_2014AR}. Each agent's kinematic model is
\begin{align*}
\dot{q}_i &= \left[ \begin{array}{c}\cos\theta_i \\\sin\theta_i\end{array}\right] \ v_i, &q_i \in \Omega, \ v_i \in \left[ 0, \ v^{\max} \right], \\
\dot{v}_i &= a_i, &a_i \in \mathbb{R},\\
\dot{\theta}_i &= \psi_i, &\theta_i \in \left[-\pi, \ \pi\right], \ \psi_i \in \mathbb{R},
\end{align*}
where $q_i$ is the agent's position, $\theta_i$ its orientation, $v_i$ its
linear velocity and $a_i, \ \psi_i$ its linear acceleration and angular
velocity control inputs. There is an upper bound $v^{\max}$ for the velocity of
the agent, while its velocity may not be negative, thus it can not move in
reverse.
It is assumed that the agent's linear velocity is due to a DC motor. By
following the method outlined in \cite{Tokekar_2014AR} and assuming the
initial and final agent velocities are zero due to the nature of the data
retrieval problem, the energy consumption of agent $i$ in the time interval
$\left[0, \ t_f\right]$ is
\begin{equation}
	\label{eq:energy}
	E_i = \int_0^{t_f} \left[c_1 a_i^2(t) + c_2 v_i^2(t) + c_3 v_i(t) + c_4 \right]
	dt,
\end{equation}
where the constants $c_1, c_2, c_3, c_4$ are dependent on the motor parameters
and $a_i$, $v_i$ denote the linear acceleration and linear velocity of agent
$i$ respectively.

\subsection{Energy Efficient Path}
Given the agent energy model \eqref{eq:energy}, an initial and final agent pose
$\left[{q_i^0}^T \ \theta_i^0 \right]^T$, $\left[{q_i^f}^T \
\theta_i^f\right]^T$ and velocities $v_i^0 = v_i^f = 0$ at times $t=0$ and
$t=t_f$ respectively, we are interested in creating an energy efficient path
from the initial to the final pose. To that extent, we use the technique
outlined in \cite{Tokekar_2014AR}, constructing a path using a sequence of
straight line and circular arc segments and finding the optimal velocity
profile along that path. In this work the derivation of optimal velocity
profiles for paths consisting of straight lines and circular segments is
presented. Additionally, an efficient algorithm for computing energy optimal paths and
their respective velocity profiles is proposed. This algorithm is also
outlined in the following paragraph.

Since optimizing both the path structure and energy consumption via the
velocity profile along the path is a difficult task, a finite number of paths
is examined instead. The agent pose $q_i, \theta_i$ and velocity $v_i$ are
discretized within some region $\Omega$. Afterwards, a directed graph
$\mathcal{G}(\mathcal{V},\mathcal{E})$ is created with the discretized
pose-velocity pairs as vertices $\mathcal{V}$ and straight line or circular arc
paths from one pose-velocity pair to another as directed edges $\mathcal{E}$.
Two vertices are connected by an edge if there exists a straight line or
circular arc path between them with a feasible velocity profile. Each graph
vertex is weighted by the energy cost of moving from its source to its target
vertex. After the graph is constructed, the $A^*$ algorithm is employed to find
the optimal path with respect to energy consumption along the graph.

\subsection{Path construction}
The action of each agent consists of a sequence of zero or more, up to
$M^{\max}$, sensor nodes so that they define route for the agent to follow.
Consequently, given a sequence of $k$ sensor nodes, the path the agent will
need to follow will consist of $k+1$ segments, given the fact that we demand
the agent returns to its original position but not necessarily its original
orientation. Since the agent must stop at each node it visits until the data
download is complete, the initial and final velocities of all segments should
be zero. Thus the construction of the total path can be simplified into the
construction of $k+1$ sub-paths. The only dependence between consecutive
sub-paths is in their initial and final poses, since the final
pose of the previous sub-path is the initial pose of the next
one.

Since only the initial orientation is known a priori, we need to define the
agent orientation at each one of the sensor nodes it will visit. Given two
consecutive sensor nodes $P_j$ and $P_{j+1}$ in an agent's path, the agent's
desired orientation at $P_j$ is arbitrarily chosen to be
\begin{equation*}
	\theta_i^j = \measuredangle P_{j+1}-P_j.
\end{equation*}
If $P_j$ is the final sensor node in the agent's path, the orientation will be
\begin{equation*}
	\theta_i^j = \measuredangle q_i^0-P_j.
\end{equation*}
Finally, the orientation of the agent when it returns at its initial position
$Q_i^0$ is the same as its orientation at the last sensor node it visited, thus
the last sub-path will be a straight line.

\section{Game Theory Preliminaries}
\label{background}
\subsection{Game Theory}
Game theory is a field of mathematics that examines how a decision should be
made when there are interactions between the parties making the decisions.
Games can be classified as either strategic form games or extensive form games.
Their principal difference is the fact that the decisions in strategic form
games are made simultaneously whereas in extensive form games decisions are
made using information about the actions of opponents.

A strategic form game consists of \cite{games1}
\begin{itemize}
\item a set of players $i \in {1,2,\ldots,I}$,
\item a set of actions $s^{i}\in S^{i}$ for each player $i$,
\item a set of joint actions, $s=(s^{1}, s^{2}, \ldots\ , s^{I}) \in S^{1} \times S^{2}\times \ldots\times S^{I}=S $,
\item the payoff function $u^{i}:S \rightarrow \mathbf{R}$ for each player $i$,
\end{itemize}
where $u^{i}(s)$ is the gain of player $i$ after the joint action $s$ has been
played. We will often denote a joint action as $s=(s^{i},s^{-i})$, where
$s^{i}$ is the action played by Player $i$ and $s^{-i}$ is the joint action
played by the opponents of Player $i$. The rules according to which the players
select the action to be played are called strategies. When the action of Player
$i$ is decided using a deterministic method, it is said that Player $i$ acts
according to a pure strategy. In case Player $i$ selects its next action based
on some probability distribution $\sigma^{i}$ over the set of available
actions, it is said that he acts according to a mixed strategy. We denote as
$\sigma=(\sigma^{1}, \ldots , \sigma^{I})$ a joint mixed strategy and will
often write $\sigma=(\sigma^{i}, \sigma^{-i})$ for simplicity, similarly to
$s=(s^{i},s^{-i})$. We denote the expected utility some player $i$ will gain by
choosing a strategy $\sigma^{i}$ (resp.\ $s^{i}$), when its opponents select
the joint strategy $\sigma^{-i}$, as $u^{i}(\sigma^{i},\sigma^{-i})$ (resp.\
$u^{i}(s^{i},s^{-i})$).

Another categorization of strategic form games is based on their payoff
functions and more precisely, they are separated into coordination and
non-coordination games.
Even though the most widely studied category are non-coordination games, they are the least
relevant to the field of distributed optimization. In coordination games
however, the goal is to maximize the gains of all players at the same time,
thus agent must cooperate in order to select a set of actions that maximizes
their common gain. Potential games are a sub-class of coordination games in
which the utilities are of the form:
\begin{equation*}
	u^{i}(s^{i},s^{-i})-u^{i}(\tilde{s^{i}},s^{-i})=
	\phi(s^{i},s^{-i})-\phi(\tilde{s^{i}},s^{-i})
\end{equation*}
where $\phi$ is a potential function. The equality holds for each player $i$,
action $s^{-i}\in S^{-i}$, and for each action pair $s^{i}$, $\tilde{s^{i}} \in S^{i}$,
where $S^{i}$ and $S^{-i}$ are the sets of all possible actions for Player $i$
and the other players respectively. A more relaxed connection between the utility
and the potential function $\phi$ is provided by ordinal potential games which
are defined as follows:
\begin{equation*}
	u^{i}(s^{i},s^{-i})-u^{i}(\tilde{s^{i}},s^{-i})>0 \Leftrightarrow
	\phi(s^{i},s^{-i})-\phi(\tilde{s^{i}},s^{-i})>0. \label{eq:poten}
\end{equation*}

Consequently, the search for an optimal solution to a distributed optimization
problem can be formulated as finding a Nash equilibrium in a potential game. A
pure Nash equilibrium \cite{Nash} is a set of pure strategies with the
additional property that when actions are chosen by the players based on these
strategies, no player can attain a better reward. It is possible to define a
Nash equilibrium as some joint mixed strategy $\hat{\sigma}$ which satisfies
the condition
\begin{equation*}
	u^{i}(\hat{\sigma}^{i},\hat{\sigma}^{-i})\geq
	u^{i}(s^{i},\hat{\sigma}^{-i}) \qquad \textrm{for all } s^{i} \in
	S^{i}
\end{equation*}
for every player $i$. It has been shown that each potential game has one or
more pure strategy Nash equilibria \cite{fp4}. We call a pure equilibrium
strict if the best reaction for each player, given the actions of its
opponents, is unique.

\subsection{Learning in games}
As described in the prequel, distributed optimization can be formulated as the
process of computing a Nash equilibrium in a game. Autonomous robots need a
coordination mechanism in order to reach their goal. This task can be
accomplished by using iterative game-theoretic learning algorithms. These
algorithms, alter the strategy of every player and thus their actions, over
time by taking into account their opponents' chosen actions. Thus, game
theoretical learning is strongly associated with iterative distributed
optimization.

One of the simplest algorithms is using the Best response decision rule and the
previously observed actions of the other players. Particularly, a player $i$
uses the previously observed actions of other players $s^{-i}_{t+1}$ and
chooses the best response to this joint action. Best response is defined as:
\begin{equation*}
	BR^{i}(\sigma^{-i})= \mathop{\rm argmax}_{s^{i} \in S^{i}} \quad
	u^{i}(s^{i},\sigma^{-i})
\end{equation*}

Another game-theoretic learning algorithm is fictitious play. Fictitious play
is possibly the most well-known learning method in game theory. The basic
concept of fictitious play is that the actions of each player are chosen as the
best response, given its expectations about the strategy of its opponent. At
first, every player has some a priori expectations about the strategy chosen by
its opponents to select their actions. At every turn, the expectations of each
player about its opponents' strategies are updated and the player acts
according to its best response. To be precise, initial arbitrary non-negative
weight functions $\kappa_{0}^{j}$, $j=1, \ldots, I$ are being updated using the
following expression:
\begin{equation}
	\kappa_{t}^{j}(s^{j}) = \kappa_{t-1}^{j}(s^{j})+I_{s^j_t=s^j}
	\label{eq:kappa}
\end{equation}
where
$I_{s^j_t=s^j}=\left\{\begin{array}{cl}1&\mbox{if
$s^j_t=s^j$}\\0&\mbox{otherwise.}\end{array}\right.$. Opponent $j$'s mixed
strategy is estimated using the expression:
\begin{equation}
	\begin{array}{l l}
	\sigma_{t}^{j}(s^{j}) &= \frac{\kappa_{t}^{j}(s^{j})}{\sum_{s^{j}\in S^{j}}\kappa_{t}^{j}(s^{j})} \\
	&= (1-\frac{1}{t})\sigma_{t-1}^{j}(s^{j})+ \frac{1}{t}I_{s^j_t=s^j} \\
	\end{array}
	\label{eq:fp}
\end{equation}

The above formula expresses the maximum likelihood estimator for the strategies
of opponents, provided that these strategies remain unchanged. Another way to
estimate opponents' mixed strategy is via the maximum a posteriori (MAP)
estimation using a Bayesian approach. Each player starts with some a priori
beliefs about the set of distributions over the opponents' actions and updates
these beliefs through the Bayes rule. If the initial assumption of the other
players' strategies is chosen to be Dirichlet distributed, then it has been
shown that the two approaches are equivalent \cite{learning_in_games}. A
different Bayesian method which goes beyond MAP estimation is presented in
\cite{leslie}.

When all players use fictitious play and a strict Nash equilibrium $\sigma$
has been played at time $t$ then it will be played for all the other iterations
of the game. Additionally any steady state of fictitious play is a Nash
equilibrium. Moreover, it has been proved that fictitious play converges for
$2\times2$ games with generic payoffs \cite{fp2}, zero sum games \cite{fp1},
games that can be solved using iterative dominance \cite{fp3} and potential
games \cite{fp4}.

Even though, fictitious play converges in various classes of games its
estimates about other players' strategies are based on expressions
(\ref{eq:kappa}) and (\ref{eq:fp}). The above updating techniques regard the
environment of the game as static and consider the probability distribution of
other players' actions to always remain unchanged, since all observations, no
matter how recent, have the same weight. Consequently, this method results in
poor response when the strategies of other players change as time progresses.

A variant which treats the most recent observations as more important than the
historic ones is geometric fictitious play \cite{learning_in_games}.
In this variant of fictitious play actions are also selected using Best
response but when forming the estimates of the other players' strategies recent
observations have higher weight while discounted the impact of historic
observations \cite{learning_in_games}. According to this variation of
fictitious play each opponent's probability to play an action $s^{j}$ is
estimated by the following formula:
\begin{equation*}
	\sigma_{t}^j(s^{j})=(1-\gamma)\sigma_{t-1}^{j}(s^{j})+ \gamma I_{s^j_t=s^j}
\end{equation*}
where $\gamma \in (0, 1) $ is a constant.

The computational time of fictitious play and geometric fictitious play
increases significantly when the number of players and their available actions
is increasing. This is because each player should estimate the strategy of
every other player. Joint strategy fictitious play is a variant of fictitious
play which was introduced in order to overcome this problem. According to the
joint strategy fictitious play algorithm each player $i$, updates in each
iteration directly its expected reward $r^{i}(s^{i})$ as follows:
\begin{equation*}
	\label{eq:jsfp_up}
	r_{t+1}^{i}(s^{i})=(1-\frac{1}{t+1})r_{t+1}^{i}(s^{i})+\frac{1}{t+1}u^{i}(s^{i}_{t},s^{-i}_{t})
\end{equation*}
The players then choose the action $s^{i}$ which has the maximum expected
reward. In order to avoid the case where more than one players change simultaneously
action and being trapped in a cycle, inertia was introduced in the joint
strategy fictitious play algorithm, Algorithm \ref{aglo:jsfpin}. Players are
choosing with probability $1$ the action which maximizes their expected reward,
$r^{i}(s^{i})$, if it coincides with the action selected in the previous
iteration. Alternatively the player choose its next action with probability:
$$\frac{\exp(r_{t+1}^{i}(s^{i}))}{ \sum_{s{j}\in
S^{i}}\exp(r_{t+1}^{i}(s^{k}))}.$$
\begin{algorithm} \caption{Joint strategy Fictitious play with inertia}
\label{aglo:jsfpin}
\begin{algorithmic}[1]
	\STATE \textit{INITIALIZATION:}\\
	\FORALL{$i \in \mathcal{I}$}
	\FORALL {$s^{i} \in S^{i}$}
	\STATE $r^{i}(s^{i})=0$
	\ENDFOR
	\STATE Uniformly choose an action.
	\ENDFOR

	\WHILE{$t< max\ iterations$}
	\FORALL{$i\in \mathcal{I}$}
	\FORALL{$s^{i}\in S^{i}$}
	\STATE$r_{t+1}^{i}(s^{i})=(1-\frac{1}{t+1})r_{t+1}^{i}(s^{i})+\frac{1}{t+1}u^{i}(s^{i}_{t},s^{-i}_{t})$
	\ENDFOR
	\ENDFOR
	\STATE $s=\mathop{\rm argmax}_{s^{i} \in S} \quad u^{i}(s^{i},\sigma^{-i})$
	\FORALL{$i\in \mathcal{I}$}
	\IF {$s^{i}_{t}=s$}
	\STATE$s^{i}_{t+1}=s$
	\ELSE
	\STATE $w~U(0,1)$
	\STATE $r_{new}=\frac{\exp(r_{t+1}^{i}(s^{i}))}{ \sum_{s{j}\in S^{i}}\exp(r_{t+1}^{i}(s^{k}))}$
	\STATE $r_{new}=sort(r_{new})$
	\FOR{$j\in \{1,2,\ldots,|S^{i}|\}$}
	\IF{ $w<\sum_{k=1:j}r_{new}(k)$}
	\STATE$s^{i}_{t}=j$
	\ENDIF
	\ENDFOR
	\ENDIF
	\ENDFOR
	\STATE{ $t=t+1$}
	\ENDWHILE
\end{algorithmic}
\end{algorithm}

\section{Utility formulation}
\label{sec:utility}
In this section a game-theoretic formulation of the MES problem is
considered. In particular the MES problem is cast as a potential game.
The provided potential game has more than one possible pure Nash equilibria.
Therefore, if autonomy is a desired property the agents (robots) need a
coordination mechanism in order to find a solution.

Consider the case where $N$ robots should visit $M$ sensors with $N<M$. The
objective is that the robots should visit all the sensors by spending the
minimum energy as a team. In addition each sensor should be visited only by a
single robot.

In a game theoretic formulation the robots are the players of the game and the
available sensors to visit represent the players' action sets. A robot can visit
up to $m$ sensors. Therefore, the case where robot $i$, $i \in \{1,\ldots,N\} $,
choose to visit sensor $j$ is a different action than the case where it decides
to visit sensors $j$ and $k$, with $j \not = k$ and $j,k \in \{1,\ldots, M\}$.
Usually when a game-theoretic formulation is used the maximization of a utility
function is considered. Therefore, the problem that players will face is the
maximization of the negative energy cost. If the energy cost for a specific
action $s^{i}$ of a player is denoted by $c^{i}$ then the following utility
function can be considered when the joint action $s=(s^{i},s^{-i})$ is played:
\begin{equation}
	r^{i}(s^{i},s^{-i})= \left \{
	\begin{array}{ c c}
	\frac{1}{c^{i}} P^{i} & \textrm{if } a^{i} \cap a^{-i}= \emptyset\\
	-c^{i} & \textrm{otherwise}
	\end{array}
	\right.
	\label{eq:reward}
\end{equation}
where $P^{i}$ is defined as $max_{s^{i}\in S^{i}}(c^{i}(s^{i}))$ and is the
energy cost of the most energy demanding action of player $i$, $a^{i}$ denote
the sensors that robot $i$ will be visiting if action $s^{i}$ is selected and
$a^{-i}$ denote the sensors that all the other robots $-i$ will be visiting if the
joint action $s^{-i}$ is selected.

The reward function in (\ref{eq:reward}) admits a potential of the form $\phi
=\sum_{i=1:m}r^{i}(s^{i},s^{-i})$. The proof of this statement is similar to
the proof of Theorem one in \cite{smyrnakis2018improving} and therefore it is omitted. The
global reward that players will share will be:
\begin{equation*}
	r(s)= \left \{
	\begin{array}{ c c}
	-C & \textrm{if } s \textrm{ is not an acceptable action}\\
	\phi & \textrm{otherwise}
	\end{array}.
	\right.
	\label{eq:reward1}
\end{equation*}
A joint action $s$ is acceptable if all the sensors of interest are visited.

\section{Simulation Studies}
\label{sec:simulations}
In this section simulation results are provided for the algorithms presented in
Section \ref{background}. In order to explore the impact of the learning
parameter $\gamma$ of geometric fictitious play in the results three instances
of geometric fictitious play were used with $\gamma=0.3, 0.1$ and $0.05$
respectively. The initial weights for all algorithms, since there was no prior
knowledge about other players actions, were chosen to be equal. The case of
three robots and four wireless sensor nodes was examined, while the energy for
each allowed action of the robots was computed according to the process
described in Section \ref{sec:model}.

In the first case robots could choose up to two sensors to visit and thus the
action space of each robot consists of seventeen actions. A robot could choose
either to take no action, or one of the four single sensors, or to choose any
combination of two sensors. In this case the best possible allocation which
provides a solution, i.e. all the target sensors are visited by only one robot,
had cost of $493.2$ Joules, and the worst 728.7.

\begin{figure}
	\centering
	\includegraphics[width=0.45\textwidth]{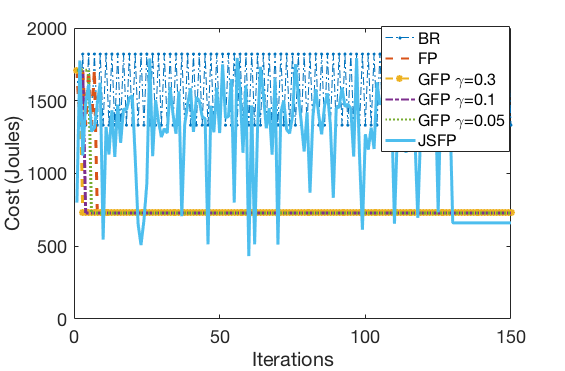}
	\caption{Team's cost for the first case.}
	\label{fig:fig2}
\end{figure}

Figure \ref{fig:fig2} depicts the results for the first case. Best Response
failed to converged to an acceptable action, when the other algorithms except
Joint strategy fictitious play converged in less than 20 iterations in a
solution. All the variants of Geometric fictitious play and the classic
fictitious play algorithm were giving as a solution the one with the worst
possible cost, 728.7 Joules. On the other hand Joint strategy fictitious play
even if it needed more iterations in order to converge to a solution, it
provided one with cost of 660.7 Joules. The relative difference between the
best possible solution and the one of Joint strategy fictitious play was
25.3\%. The relative change is defined as: $$\Delta=\frac{|x-y|}{\max(x,y)}.$$.

In the second case robots could visit up to three sensors. Thus, the action space
has been increased to 41 possible actions per robot. In addition to the
actions, available to each robot in the second case, the action space includes
all the possible combinations of visiting three sensors.

\begin{figure}
	\centering
	\includegraphics[width=0.45\textwidth]{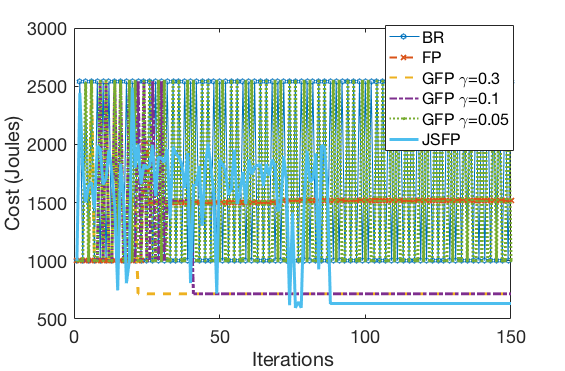}
	\caption{Team's cost for the third case.}
	\label{fig:fig3}
\end{figure}

As it is depicted in Figure \ref{fig:fig3} GFP with $\gamma=0.05$, Fictitious
play and Best response fails to converge to a solution. The remaining two
instances of Geometric fictitious play converge to the same solution with cost
$716.4$ Joules, which is 31.16\% worst than the best possible solution. Joint
strategy fictitious play needed more iterations to converge to a solution but
its corresponding cost was smaller than the two Geometric Fictitious play
instances. In particular its cost was 633.3 Joules, 22.12\% worst than the most
energy efficient solution.

\begin{figure}
	\centering
	\includegraphics[width=0.4\textwidth]{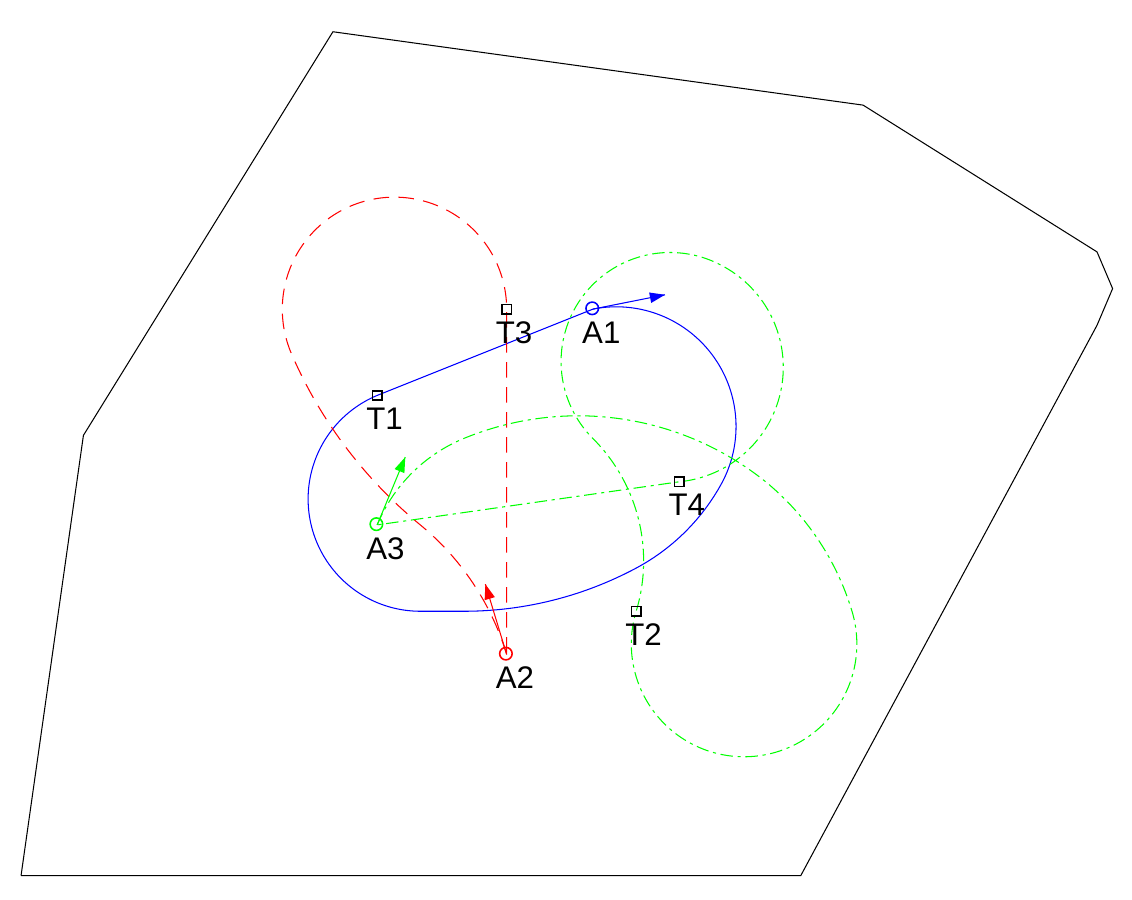}
	\caption{Trajectories of the robots for the first case when Fictitious play algorithm is used.}
	\label{fig:traj1}
\end{figure}

\begin{figure}
	\centering
	\includegraphics[width=0.4\textwidth]{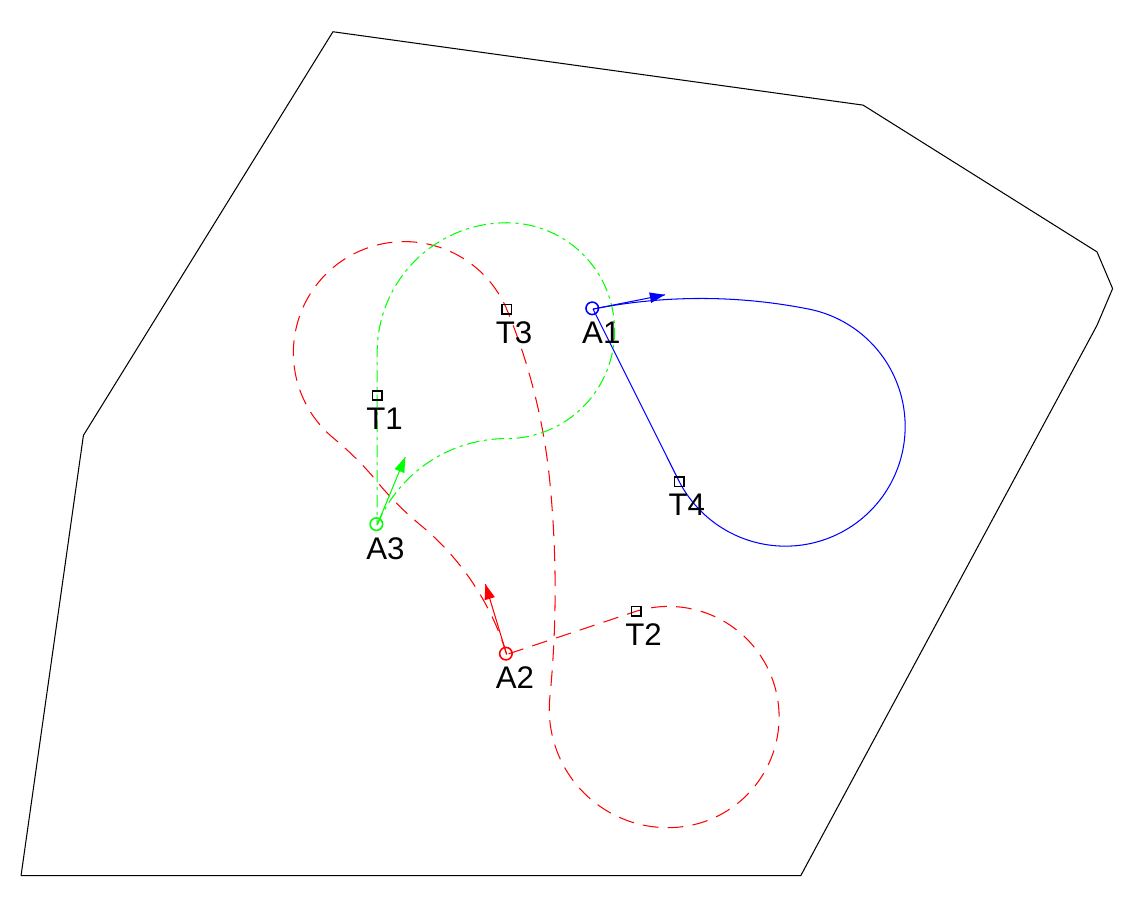}
	\caption{Trajectories of the robots for the second case when Joint strategy fictitious play algorithm is used.}
	\label{fig:traj2}
\end{figure}

Samples of the trajectories selected by robots for the two cases are depicted
in Figures \ref{fig:traj1} and \ref{fig:traj2}. The robots are denoted A$1$
through A$3$ and their initial poses are marked as arrows, while the targets
are denoted T$1$ through T$4$ and their positions are marked with squares.

It should be reported that a typical tenfold reduction of time has been
observed compared to the classical MILP solver.

\section{Conclusions}
A game-theoretic formulation for the energy efficient data retrieval problem
from stationary nodes using mobile agents is presented. In particular, it is
shown that this problem can be cast as a potential game with more than one Nash
equilibria. Therefore, a coordination mechanism is needed in order for the
robots to reach an acceptable solution. For that reason four basic learning
algorithms from game-theoretic literature were used. Even though these
algorithms are trapped in local optimum solutions, their computational cost
allow them to be utilized in real time applications.

\bibliographystyle{IEEEtran}
\bibliography{references}
\end{document}